\documentclass[preprint,review,12pt]{elsarticle} 
\usepackage{booktabs} 
\usepackage{siunitx} 
\usepackage{pgfplotstable} 
\usepackage[latin1]{inputenc}
\usepackage{verbatim}
\usepackage{graphicx}
\usepackage{supertabular}
\usepackage{color}
\usepackage{amsmath}
\usepackage{amsfonts}
\usepackage{amssymb}
\usepackage{subfigure}
\usepackage{vmargin}
\usepackage[numbers]{natbib}
\usepackage{wrapfig}
\usepackage{multirow} 
\usepackage{wasysym}
\usepackage{array}
\usepackage{tikz}
\usepackage{csvsimple}
\usetikzlibrary{shapes.geometric}
\usetikzlibrary{patterns}

\begin{document}
\graphicspath{{img/}}
\begin{frontmatter}
\title{Experimental \textit{n}-Hexane-Air Expanding Spherical Flames}
\author{Stephanie A. Coronel$^{\mathrm{a,}}$\corref{cor1}} 
\author{Simon Lapointe$^{\mathrm{a,}}$\corref{cor2}} 
\author{R\'emy M\'evel$^{\mathrm{b,c}}$\corref{cor3}}
\author{Vaughan  L. Thomas$^{\mathrm{d}}$\corref{cor4}}
\author{Nabiha Chaumeix$^{\mathrm{e}}$\corref{cor6}}
\author{Joseph E. Shepherd$^{\mathrm{a}}$\corref{cor5}}
\cortext[cor1]{Corresponding author: scorone@sandia.gov; current address: Sandia National Laboratories, Albuquerque, NM}
\cortext[cor2]{Current address: Lawrence Livermore National Laboratories, Livermore, CA}
\address{$\mathrm{^a}$Graduate Aerospace Laboratories, California Institute of Technology, Pasadena, California 91125, USA\\
$\mathrm{^b}$Center for Combustion Energy, $\mathrm{^c}$Department of Automotive Engineering, Tsinghua University, Beijing, China\\
$\mathrm{^d}$Department of Mechanical Engineering, Johns Hopkins University, Baltimore, Maryland 21218, USA\\
$\mathrm{^e}$ICARE, CNRS-INSIS, 1C Avenue de la Recherche Scientifique, 45071 Orl\'eans Cedex 2, France\\}
\begin{abstract}
The effects of initial pressure and temperature on the laminar burning speed of \textit{n}-hexane-air mixtures were investigated experimentally and numerically. The spherically expanding flame technique with a nonlinear extrapolation procedure was employed to measure the laminar burning speed at atmospheric and sub-atmospheric pressures and at nominal temperatures ranging from 296 to 422 K. The results indicated that the laminar burning speed increases as pressure decreases and as temperature increases. The predictions of three reaction models taken from the literature were compared with the experimental results from the present study  and previous data for \textit{n}-hexane-air mixtures. Based on a quantitative analysis of the model performances, it was found that the most appropriate model to use for predicting laminar flame properties of \textit{n}-hexane-air mixtures is JetSurF.
\end{abstract}
\end{frontmatter}

\clearpage

\section{Introduction}

During aircraft operation, the pressure within the fuel tank and other areas potentially containing flammable mixtures varies between 20 and 100 kPa. To assess the risk of potential ignition hazards and flammability in fuel tank ullage or flammable leakage zones, it is necessary to characterize properties such as the laminar burning rate of fuel-air mixtures over a wide range of initial pressures and temperatures. \textit{n}-Hexane has been extensively used at the Explosion Dynamics Laboratory as a single component surrogate of kerosene \citep{HeatingRate,SallyThesis,PhilThesis,Menon2016}; \textit{n}-hexane exhibits a relatively high vapor pressure which facilitates experimenting at ambient temperature. In contrast to \textit{n}-heptane, which has been widely studied, \textit{n}-hexane oxidation has received little interest \citep{SimmieReview}. \citet{CurranHexane} studied hexane isomer chemistry through the measurement and modeling of exhaust gases from an engine. The ignition delay-time behind a shock wave was measured by \citet{BurcatHexane,StarikovskiiHexane,Zhang2015,Mevel2016}. \citet{Zhang2015} also measured the ignition delay-time in the low-temperature regime using a rapid compression machine as well as species profiles using the jet-stirred reactor technique. \citet{MevelFuel2014} employed a flow reactor along with gas chromatography (GC) analyses and laser-based diagnostics to measure the species profiles in the temperature range $600-1000$ K. \citet{HeatingRate} studied the effect of the heating rate on the low temperature oxidation of \textit{n}-hexane by air, and the minimum temperature of a heated surface required to ignite \textit{n}-hexane-air mixtures \cite{Menon2016}. \citet{SallyThesis} measured the minimum ignition energy of several \textit{n}-hexane-air mixtures. A limited number of studies have been found on the laminar burning speed. \citet{DavisLaw} measured the laminar burning speed of \textit{n}-hexane-air mixtures at ambient conditions using the counterflow twin flame technique. \citet{FarellHexane} used pressure traces from spherically expanding flames to determine the laminar burning speed of \textit{n}-hexane-air mixtures at an initial temperature and pressure of $450$ K and 304 kPa, respectively. \citet{LawHexane} reported experimental measurements using spherically expanding flames at an initial temperature of $353$ K and an initial pressure range of $100-1000$ kPa. \citet{JiHexaneFlame} used the counterflow burner technique to measure the laminar burning speed of \textit{n}-hexane-air mixtures at an initial temperature and pressure of $353$ K and $100$ kPa, respectively.

In contrast to previous work, the present study focuses on initial conditions below atmospheric pressure in order to simulate aircraft fuel tank conditions. Additionally, this study investigates the effect of initial temperature at sub-atmospheric conditions to simulate elevated temperature conditions in the fuel tank ullage or flammable leakage zones.
\section{Experimental Setup and Methodology}\label{sec:experiment}
\subsection{Facilities}
Two experimental facilities were used in the present study to cover a wide range of initial temperature conditions: the Explosion Dynamics Laboratory (EDL) at the California Institute of Technology (Caltech) and the Institut de Combustion A\'erothermique R\'eactivit\'e et Environnement (ICARE)-Centre National de la Recherche Scientifique (CNRS) Orl\'eans. At the EDL, the experiments were performed in a $22$ L stainless steel combustion vessel. Parallel flanges were used to mount electrodes for the ignition system and windows for optical access. The mixtures were ignited by a 300 mJ electric spark generated between two $0.4$ mm in diameter tungsten electrodes separated by a distance of $2-4$ mm. A high-speed camera (Phantom v711) was used to record the flame propagation observed using Schlieren visualization and shadowgraphy at a rate of $10,000$ frames per second with a resolution of $512\times512$ px. The experiments conducted at ICARE-CNRS were performed in a stainless steel spherical bomb consisting of two concentric spheres; the internal sphere had an inner diameter of $476$ mm. The mixtures were ignited by electric sparks with a nominal energy of $1.82$ mJ. Schlieren visualization was used with a high-speed camera (Phantom V1610) at a rate of $25,000$ frames per second with a resolution of $768\times768$ px.

\subsection{Flame Edge Detection}
The flame radius as a function of time was extracted from the experimental images of expanding spherical flames using algorithms implemented in Matlab, including an edge detection operator \cite{Nativel20161,Mevel2009}.
The images of the spherically propagating flames were processed by first applying a mask over each image to remove the background (electrodes). Edge detection was then used to identify the expanding flame edge. An ellipse was fitted to the detected flame edge; the ellipse parameters were then used to obtain an equivalent radius. For the majority of the experimental images, the flame sphericity was approximately equal to 1.

\subsection{Extrapolation of Flame Parameters}\label{sec:nonlinear}
Using asymptotic methods based on large activation energy, \citet{SivashinskyTheory} obtained a nonlinear model for spherical flame speed as a function of curvature (Eq.~\ref{RSequation}).
\begin{equation}
\left(\frac{S_b}{S^{0}_{b}}\right)^{2} \ln \left(\frac{S_b}{S^{0}_{b}}\right)^{2}  =-2 \frac{L_{B}  \kappa}{S^{0}_{b}}.
\label{RSequation}
\end{equation}
$S_b$ and $S_b^0$ are the stretched and unstretched flame speeds, respectively, $L_B$ is the burnt gas Markstein length, and $\kappa$ is the stretch rate. 
\citet{Karlovitz} expressed the stretch rate in terms of the normalized rate of change of an elementary flame front area as,
\begin{equation}
\kappa=\frac{1}{A} \frac{\text{d}A}{\text{d}t} ,
\label{Karlovitz}
\end{equation}
where $A$ is the flame front area. In the case of a spherical flame, the flame surface is given by $A=4 \pi R_{f}^{2}$, leading to the following expression for the stretch rate \citep{Lamoureux2,Aung1,Dowdy,Jerzembeck}:
\begin{equation}
\kappa=2 \frac{S_{b}}{R_{f}} ,
\label{StretchRate}
\end{equation}
and given that the flame speed corresponds to the flame radius increase rate,
\begin{equation}
S_{b}=\frac{\text{d}R_{f}}{\text{d}t}. 
\label{StretchRateBis}
\end{equation}
The measured rate of increase of the flame radius, $\text{d}R_f/\text{d}t$, is assumed to be the flame speed  since the combustion products are stationary in the laboratory frame. In the case of a large volume vessel and for measurements limited to the initial period of propagation when the flame radius is small compared to the experimental set-up dimensions, the pressure increase can be neglected \citep{BradleyEvolutionPression}.

Combining Eqs.~\ref{StretchRate} and \ref{RSequation} and simplifying the logarithmic term leads to the following relation,
\begin{equation}
\frac{S_{b}}{S^{0}_{b}} \ln\left(\frac{S_{b}}{S^{0}_{b}}\right) =- 2\frac{L_{B}}{R_f}  \ .
\label{Propagation_IDE_Simplified_3}
\end{equation}  
Since the flame speed is positive, the term on the left hand side may take on values only within the range $[-e^{-1},\infty)$. For $L_{B}<0$ a solution exists for all positive values of $R_f$, but for $L_{B}>0$, a solution exists only if ,
\begin{align}
\frac{R_f}{2  L_{B}} &\ge e \qquad\qquad (L_{B}>0).  \label{RS_limit}
\end{align}
Thus for positive Markstein lengths, there exists a minimum flame radius below which the quasi-steady relationship between flame speed and stretch rate is not valid, and hence the unstretched flame speed cannot be extracted using Eqs.~\ref{RSequation} or \ref{Propagation_IDE_Simplified_3}. This constraint can be viewed as a maximum Markstein length, $L_{B,\text{max}}$, for a fixed minimum (or initial) flame radius. The fact that no solutions exist for small flame radii is a consequence of the neglected unsteady term which is important in the early-time flame dynamics \citep{SivashinskyTheory}. This limitation was also identified by \citet{Lipatnikov2015}.

Equation~\ref{Propagation_IDE_Simplified_3} is used to derive the unstretched flame speed and the Markstein length from experimental data. One approach to doing this is to analyze the flame radius history $R_{f}=f(t)$ data applying polynomial fits and differentiating to determine $S_b=\text{d}R_f/\text{d}t$ \citep{HalterNonLinear,BouvetSyngas}. Numerical differentiation of the experimental data leads to amplification of existing noise. To avoid differentiating  the experimental data, \citet{KelleyLawNonlinear} proposed an integrated form of Eq.~\ref{RSequation}. In the present study, numerical integration rather than analytic integration is used to extract the flame properties from the nonlinear result of \citet{SivashinskyTheory}. The unstretched burning speed, $S_u^0$ is obtained through $S_{u}^0=S_{b}^0/\sigma$, where $\sigma$ is the expansion ratio defined as $\sigma=\rho_{u}/\rho_{b}$, where $\rho_{u}$ and $\rho_{b}$ are the unburnt and burnt gas densities, respectively. For the remainder of this study, the unstretched burning speed will be referred to as the laminar burning speed.

\section{Results and Discussion}
\subsection{Experimental Results}

\begin{figure}[h!]
\begin{center}
\includegraphics[scale=0.8]{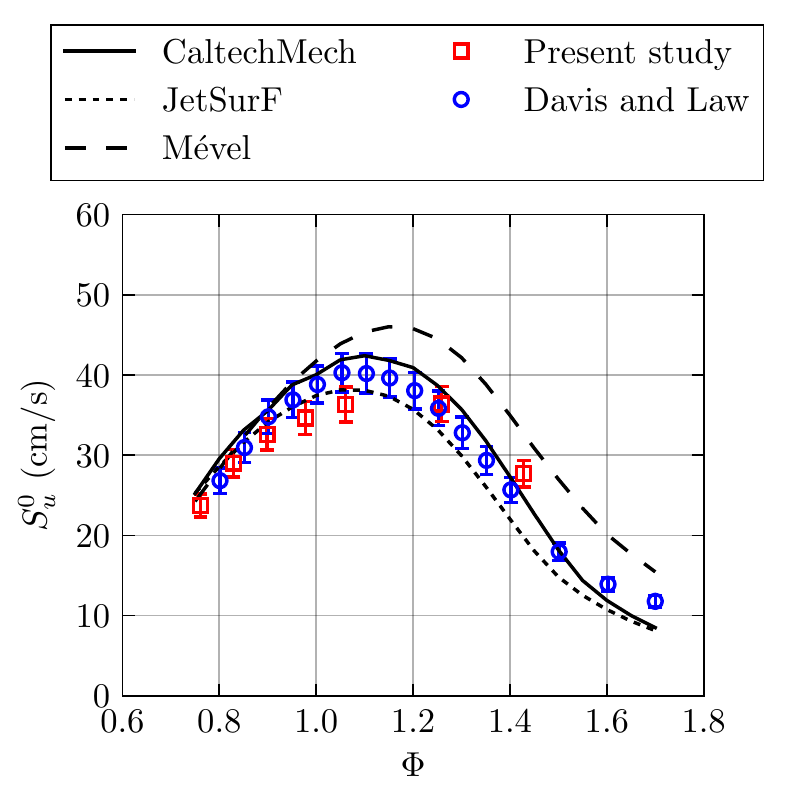}
\caption{Experimental laminar burning speed of \textit{n}-hexane-air mixtures as a function of equivalence ratio at a nominal initial temperature and pressure of 300 K and 100 kPa, respectively, along with numerical calculations (CaltechMech \cite{BlanquartMech}, JetSurF \cite{jetsurf}, and M\'evel \cite{MevelFuel2014}).}
\label{fig:Experimental100kPa}      
\end{center}
\end{figure}

Experimental laminar burning speeds at an initial temperature of 296 K and pressure  of 100 kPa are shown in Fig.~\ref{fig:Experimental100kPa} along with results previously obtained by \citet{DavisLaw}. The uncertainty in the laminar burning speeds is on average $6\%$, the value is based on previous estimates made by \citet{Mevel2009} who used the same flame detection algorithms employed in the present study. Figure~\ref{fig:Experimental100kPa} also shows 1D freely propagating flame calculations performed using FlameMaster \citep{pitsch} with three different chemical kinetic mechanisms: CaltechMech \cite{BlanquartMech}, JetSurF \cite{jetsurf}, and the mechanism of \citet{MevelFuel2014} (referred to as M\'evel in this study). Further details on mechanism description and performance are provided in Section~\ref{sec:model}. A Mann-Whitney-Wilcoxon (MWW) RankSum test indicated that the differences in the two laminar burning speed distributions shown in Fig.~\ref{fig:Experimental100kPa}  were not statistically significant; details of the test can be found in the Appendix. 

\begin{figure}[h!]
\begin{center}
\includegraphics[scale=0.8]{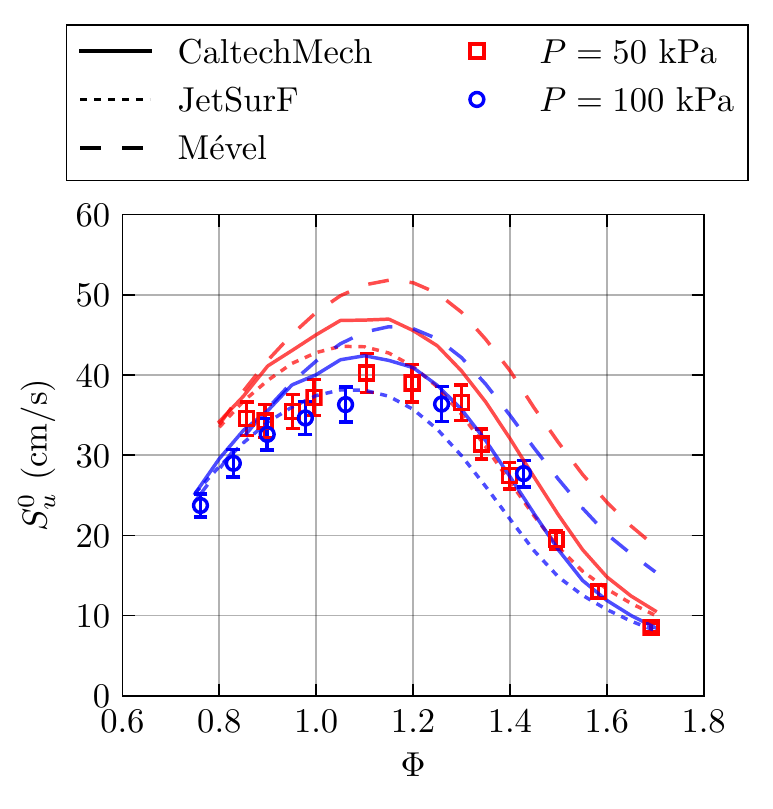}
\caption{Experimental laminar burning speed of \textit{n}-hexane-air mixtures as a function of equivalence ratio at nominal initial pressures of 50 kPa and 100 kPa and nominal initial temperature of 300 K; numerical calculations (CaltechMech \cite{BlanquartMech}, JetSurF \cite{jetsurf}, and M\'evel \cite{MevelFuel2014}) also shown.}
\label{fig:Experimental50kPa100kPa}    
\end{center}
\end{figure}

The evolution of the laminar burning speed  as a function of equivalence ratio was studied at a nominal initial temperature and pressure of $300$ K and $50$ kPa, respectively. Figure~\ref{fig:Experimental50kPa100kPa} shows the laminar burning speed obtained at initial pressures of $100$ kPa and $50$ kPa. The MWW RankSum test indicated that the differences in the laminar burning speed distributions at 100 kPa and 50 kPa were not statistically significant.  

\begin{figure}[h!]
\begin{center}
\includegraphics[scale=0.8]{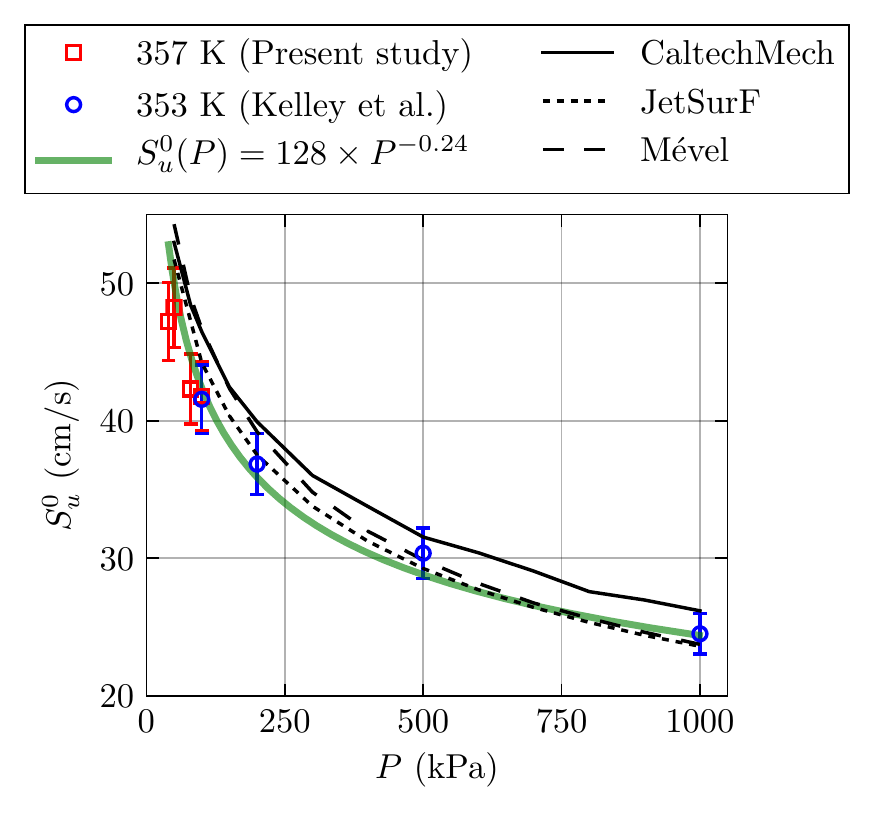} \\
\caption{Experimental laminar burning speed  of \textit{n}-hexane-air mixtures as a function of initial pressure at an initial temperature of 353 and 357 K along with numerical calculations (CaltechMech \cite{BlanquartMech}, JetSurF \cite{jetsurf}, and M\'evel \cite{MevelFuel2014}).}
\label{fig:ExperimentalP} 
\end{center}
\end{figure}

The effect of initial pressure on the laminar burning speed was investigated at $\Phi=0.90$ and a nominal initial temperature of $357$ K. The experimental laminar burning speed is shown in Fig.~\ref{fig:ExperimentalP} along with experimental results obtained by \citet{LawHexane} at initial pressures of $100-1000$ kPa and an initial temperature of 353 K. The laminar burning speed decreases with increasing initial pressure, $20\%$ between $50$ and $100$ kPa and $53\%$ between $50$ and $1000$ kPa at nominal initial temperatures of $353$ and 357 K. The pressure dependence on the laminar burning speed can be fit to a power law: $ S_u^0\left(P\right) = 128 \times P^{-0.24}$, where $P$ has units of kPa. The corresponding standard deviations for the pre-exponential and exponent are 12 and 0.02, respectively.

\begin{figure}[h!]
\begin{center}
\includegraphics[scale=0.8]{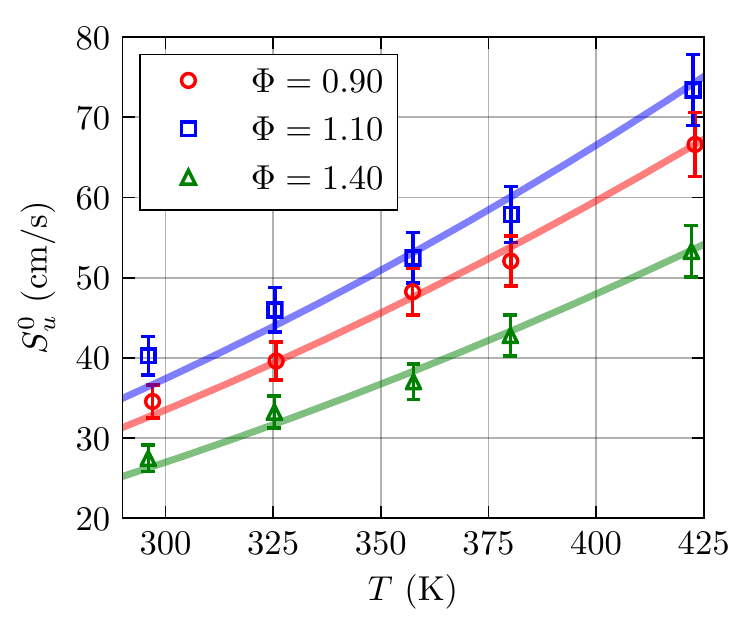}
\caption{Experimental laminar burning speed of \textit{n}-hexane-air mixtures as a function of initial temperature at an initial pressure of 50 kPa and $\Phi=0.9,\,1.1,\,\text{and}\,1.4$; the solid lines correspond to $S_u^0\sim T^2$.}
\label{fig:ExperimentalT}    
\end{center}
\end{figure}

The effect of initial temperature was studied at an initial pressure of $50$ kPa and three equivalence ratios, $\Phi=\{0.90,1.10,1.40\}$. The laminar burning speed and flux are shown in Fig.~\ref{fig:ExperimentalT}. At initial temperatures of $296$ K to $422$ K, the laminar burning speed increases by approximately $93\%$, $82\%$, and $94\%$ for $\Phi=0.90$, $\Phi=1.10$, and $\Phi=1.40$, respectively. There is a distinct difference between the laminar burning speeds distributions shown for $\Phi=\{0.90,1.10,1.40\}$. Each distribution can be fit to a power law $S_u^0 \sim T^2$ shown in Fig.~\ref{fig:ExperimentalT}; however, the best fit for each distribution is  $S_u^0 \sim T^{1.9}$ ($\Phi=0.90$), $S_u^0 \sim T^{1.7}$ ($\Phi=1.10$), and $S_u^0 \sim T^{1.9}$ ($\Phi=1.40$). The standard deviation of the exponents in the best fits  is 0.1.

\begin{figure}[h!]
\begin{center}
\includegraphics[scale=0.8]{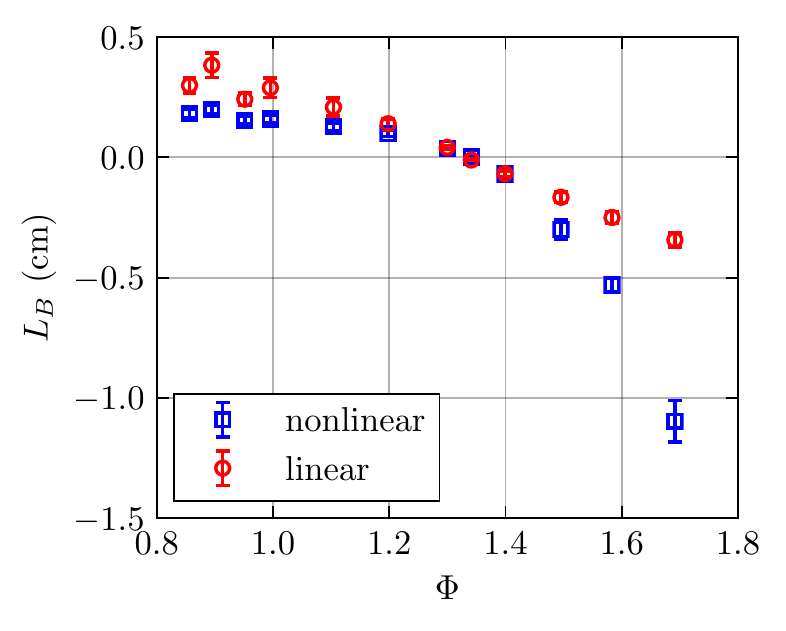}
\caption{Evolution of the Markstein length for \textit{n}-hexane-air mixtures as a function of equivalence ratio at a nominal initial temperature and pressure of 296 K and 50 kPa, respectively, using linear and nonlinear extrapolations.}
\label{fig:Markstein}      
\end{center}
\end{figure}

Figure~\ref{fig:Markstein} shows the variation of the Markstein length with equivalence ratio at an initial temperature and pressure of $296$ K and $50$ kPa, respectively. Lean and rich mixtures exhibit positive and negative Markstein lengths, respectively. The transition from positive to negative Markstein length occurs at $\Phi \approx 1.3$. This trend is consistent with previous Markstein length results obtained for C$_{5}$ to C$_{8}$ \textit{n}-alkane-air mixtures \citep{LawHexane}. Figure~\ref{fig:Markstein}  shows the Markstein length extrapolated using a linear and nonlinear dependence of the stretched flame speed on stretch rate. The linear dependence on stretch rate is given by $S_b=S_b^0-L_B\kappa$. It is evident from the figure that deviations of the nonlinear $L_B$ from the linear $L_B$ occur for both rich and lean \textit{n}-hexane-air mixtures.

The radii range and number of points used to extract the Markstein lengths of Fig.~\ref{fig:Markstein} are shown in Table~\ref{table:conditions} where N is the number of flame radius points, and $R_{f_0}$ and $R_{f_N}$ are the initial and final flame radius. The values of $R_{f_N}$ across all tests is between 40 and 50 cm; \citet{HUO2018} indicated that a final flame radius of 40 cm compared to 20 cm reduced the error in extrapolation of the flame parameters from $6\%$ to $3\%$ and $8\%$ to $4\%$ for H$_2$-air at $\Phi=4.5$ and C$_3$H$_8$-air at $\Phi=0.8$, respectively.
\begin{table}[ht!]
\begin{center}
\begin{tabular}{cccccc}
\hline
Test &	$\Phi$ &	N &	Range (mm) &	$R_{f_0}$ (mm) & $R_{f_N}$ (mm)\\
\hline
24 &	0.85 &	 147 &	32 &	14 &		46\\
44 &	0.86 &	139 &	30 &	14 &		44\\
20 &	0.89 &	168 &	34 &	12 &		46\\
40 &	0.90 &	159 &	36 &	11 &		47\\
43 &	0.95 &	119 &	31 &	14 &		45\\
26 &	0.99 &	160 &	37 &	10 &		47\\
18 &	1.00 &	149 &	39 &	9 &	 	48\\
27 &	1.10 &	129 &	36 &	9 &		 47\\
39 &	1.11 &	124 &	37 &	10 &		47\\
29 &	1.20 &	128 &	37 &	9 &	 	46\\
30 &	1.20 &	123 &	36 &	10 &	 	46\\
9 & 1.30 &		116 &	36 &	8 &	 	44\\
31 &	1.30 &	139 &	37 &	10 &	 	47\\
41 &	1.34 &	140 &	35 &	10 &	 	45\\
32 &	1.40 &	155 &	35 &	10 & 		45\\
33 &	1.50 &	193 &	35 &	10 & 		45\\
34 &	1.58 &	166 &	20 &	25 &		45\\
42 &	1.69 &	219 &	22 &	19 &		41\\
\hline
\end{tabular}
\end{center}
\caption{Properties of experimental flame radius distributions used in obtaining Markstein lengths shown in Fig.~\ref{fig:Markstein}.}
\label{table:conditions}
\end{table}

\begin{figure}[h!]
\begin{center}
\includegraphics[scale=0.8]{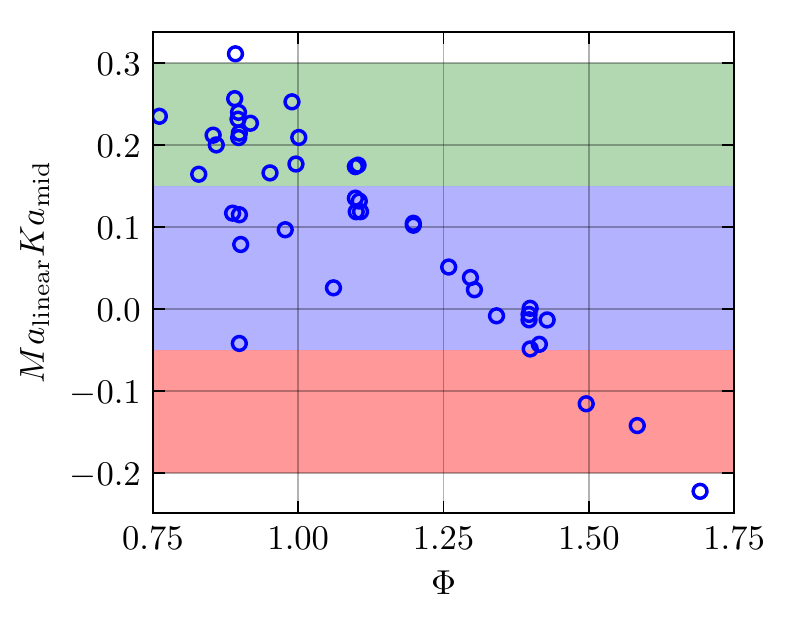}
\caption{$Ma_{\mathrm{linear}}Ka_{\mathrm{mid}}$ as a function of equivalence ratio for initial temperatures and pressures of  296 K to 380 K, and 40 kPa to 100 kPa, respectively.}
\label{fig:MaKa}      
\end{center}
\end{figure}

Figure~\ref{fig:MaKa} shows the product of the Markstein number, $Ma_{\mathrm{linear}}$ (obtained via the linear extrapolation method), and the Karlovitz number, $Ka_{\mathrm{mid}}$ (evaluated at the mid-point of the flame radii data), as a function of the mixture equivalence ratio. The product is suggested by \citet{Wu2015} as a method to evaluate the uncertainty of the extrapolation method. In Fig.~\ref{fig:MaKa}, the blue, green, and red regions have extrapolation uncertainties of $\le5\%$, $5-12\%$, and $5-40\%$, respectively. The points lying in the red region correspond to rich conditions at a nominal initial temperature and pressure of 296 K and 50 kPa, respectively.

\begin{figure}[h!]
\begin{center}
\includegraphics[scale=1]{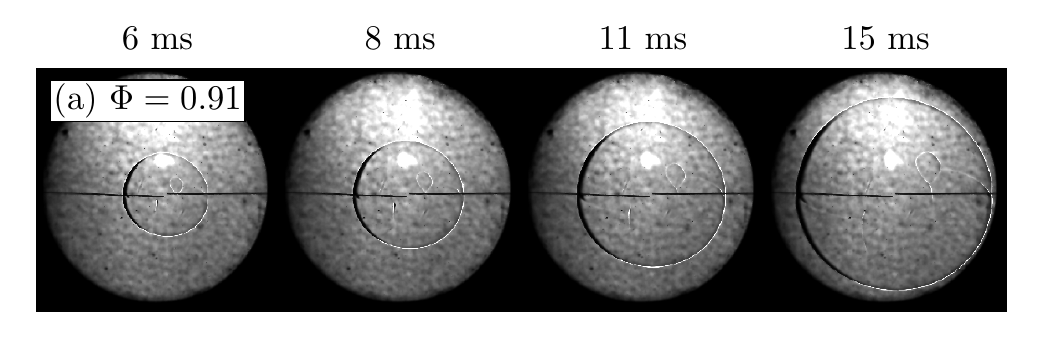}\\
\vspace{0.2cm}
\includegraphics[scale=1]{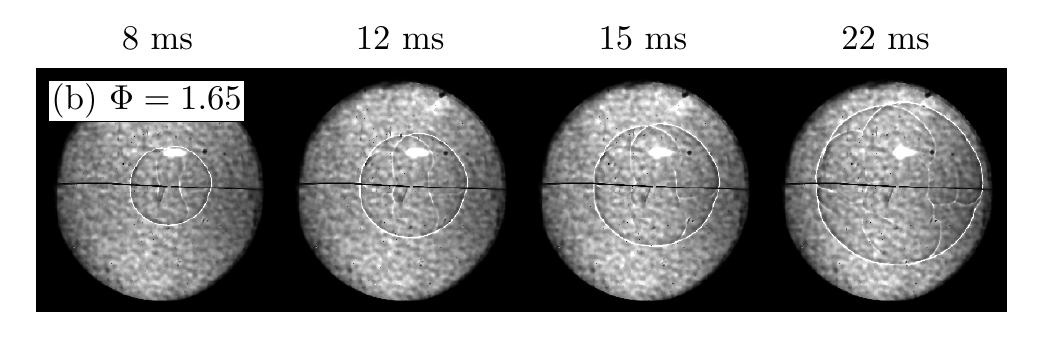} \\
\caption{Example of (a) stable and (b) unstable flame propagation of \textit{n}-hexane-air mixtures at an initial temperature and initial pressure of 296 K and 50 kPa, respectively.}
\label{MarksteinPropagation}      
\end{center}
\end{figure}

Figure~\ref{MarksteinPropagation} shows examples of a stable lean mixture and an unstable rich mixture flame propagation. For  the  lean mixture shown in Fig.~\ref{MarksteinPropagation} (a), the flame front remains smooth and undisturbed during the propagation within the field of view $R_{f} \leq R_{\text{window}}$, where $R_\text{window}$ is the window radius. For the rich mixture shown in Fig.~\ref{MarksteinPropagation} (b), the flame front becomes progressively more disturbed as it grows, and exhibits significant cellular structures before the flame exits the field of view. The development of the cellular pattern is likely due to thermo-diffusive instabilities that are characteristic of rich hydrocarbon-air mixtures \citep{LawCell1}. These instabilities create a flame that is no longer spherical and therefore the flame radius measurements are no longer correct because of the unknown relationship between the average flame radius and the flame surface.

\subsection{Modeling Results}\label{sec:model}
The 1D freely propagating flame calculations performed with FlameMaster  \citep{pitsch} used the chemical kinetic mechanisms of CaltechMech \cite{BlanquartMech}, JetSurF \cite{jetsurf}, and M\'evel \cite{MevelFuel2014}. The calculations neglected Soret and Dufour effects, and a mixture-averaged formulation was used for the transport properties. \citet{JiHexaneFlame} showed that using a multicomponent transport coefficient formulation rather than mixture-averaged transport properties resulted in a 1 cm/s increase in the calculated laminar burning speeds of C$_5$-C$_{12}$ \textit{n}-alkane mixtures. A study by \citet{xin2012} found that accounting for Soret effects resulted in a maximum of  $1-2\%$ increase in the laminar burning speed of \textit{n}-heptane-air flames at and near stoichiometric conditions. Finally, \citet{bongers2003} showed that for C$_3$ laminar premixed flames, the effect of excluding Dufour effects was negligible. 

\citet{BlanquartMech} developed CaltechMech for the combustion of engine relevant fuels; the mechanism consists of 172 species and 1,119 reactions. It should be noted that \citet{BlanquartMech} placed importance on the accurate modeling of formation of soot precursors for fuel surrogates in premixed and diffusion flames.  \citet{BlanquartMech} performed extensive validation of CaltechMech  using experimental ignition delay time and laminar burning speed data. The flame calculations performed by \citet{BlanquartMech} included Soret and Dufour effects, and mixture-averaged transport properties.

\begin{figure}[h!]
\begin{center}
\includegraphics[scale=0.8]{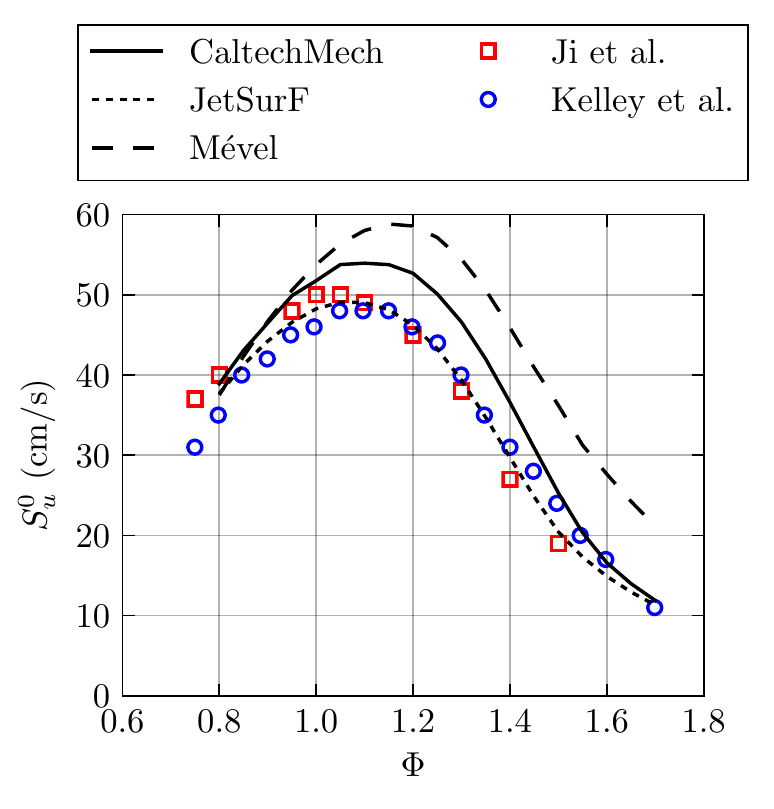}
\caption{Experimental and numerical (CaltechMech \cite{BlanquartMech}, JetSurF \cite{jetsurf}, and M\'evel \cite{MevelFuel2014}) laminar burning speed of \textit{n}-hexane-air mixtures as a function of equivalence ratio at an initial temperature and pressure of $353$ K and $100$ kPa, respectively.}
\label{fig:T353K}    
\end{center}
\end{figure}

\citet{jetsurf} developed JetSurF for high temperature applications of \textit{n}-alkanes, along with other fuels (cyclohexane, and methyl-,ethyl-,\textit{n}-propyl and \textit{n}-butyl-cyclohexane). The JetSurF version used in the present study consists of 348 species and 2,163 reactions. Calculations have been performed with previous versions of JetSurF and compared against experimental laminar burning speeds of \textit{n}-alkanes  by \citet{DavisLaw,you2009,smallbone2009,JiHexaneFlame,LawHexane}. Experimental laminar burning speed measurements used for comparison with JetSurF 1.0 calculations were performed by \citet{JiHexaneFlame,LawHexane}; the results are shown in Fig.~\ref{fig:T353K} along with the modeling results obtained in the present study.

\citet{MevelFuel2014} developed the last chemical kinetic reaction mechanism, consisting of 531 species and 2,628 reactions,  presented in this study. The mechanism was not validated against experimental laminar burning speeds since that was outside the scope of the study presented by \citet{MevelFuel2014}.

\begin{figure}[h!]
\begin{center}
\includegraphics[scale=0.8]{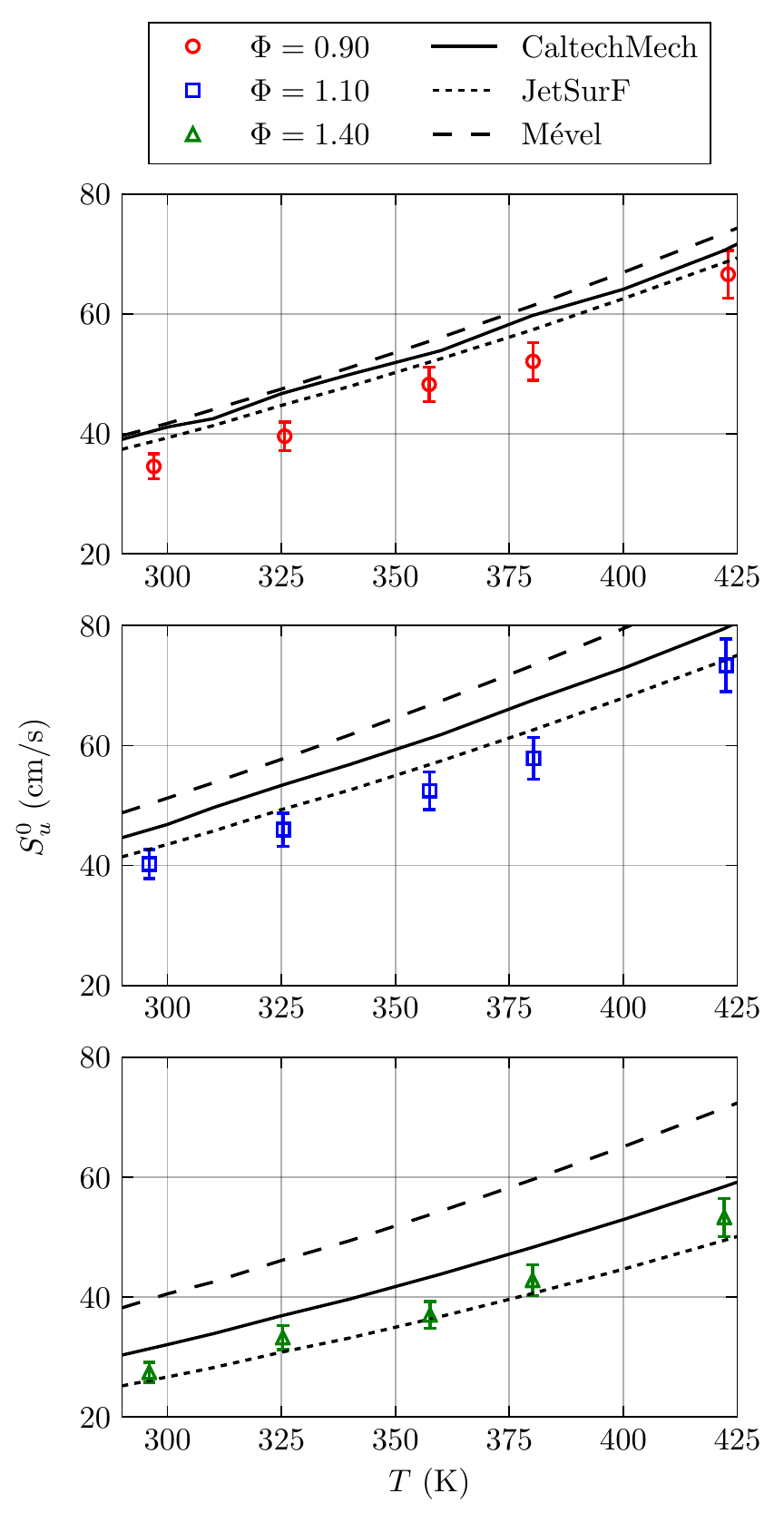}
\caption{Experimental laminar burning speed of \textit{n}-hexane-air mixtures as a function of initial temperature at an initial pressure of 50 kPa and $\Phi=0.9,\,1.1,\,\text{and}\,1.4$ along with numerical calculations (CaltechMech \cite{BlanquartMech}, JetSurF \cite{jetsurf}, and M\'evel \cite{MevelFuel2014}).}
\label{fig:Tmodel}    
\end{center}
\end{figure}

\subsubsection{Model Performance}
\begin{figure}[ht!]
\begin{center}
\includegraphics[scale=0.8]{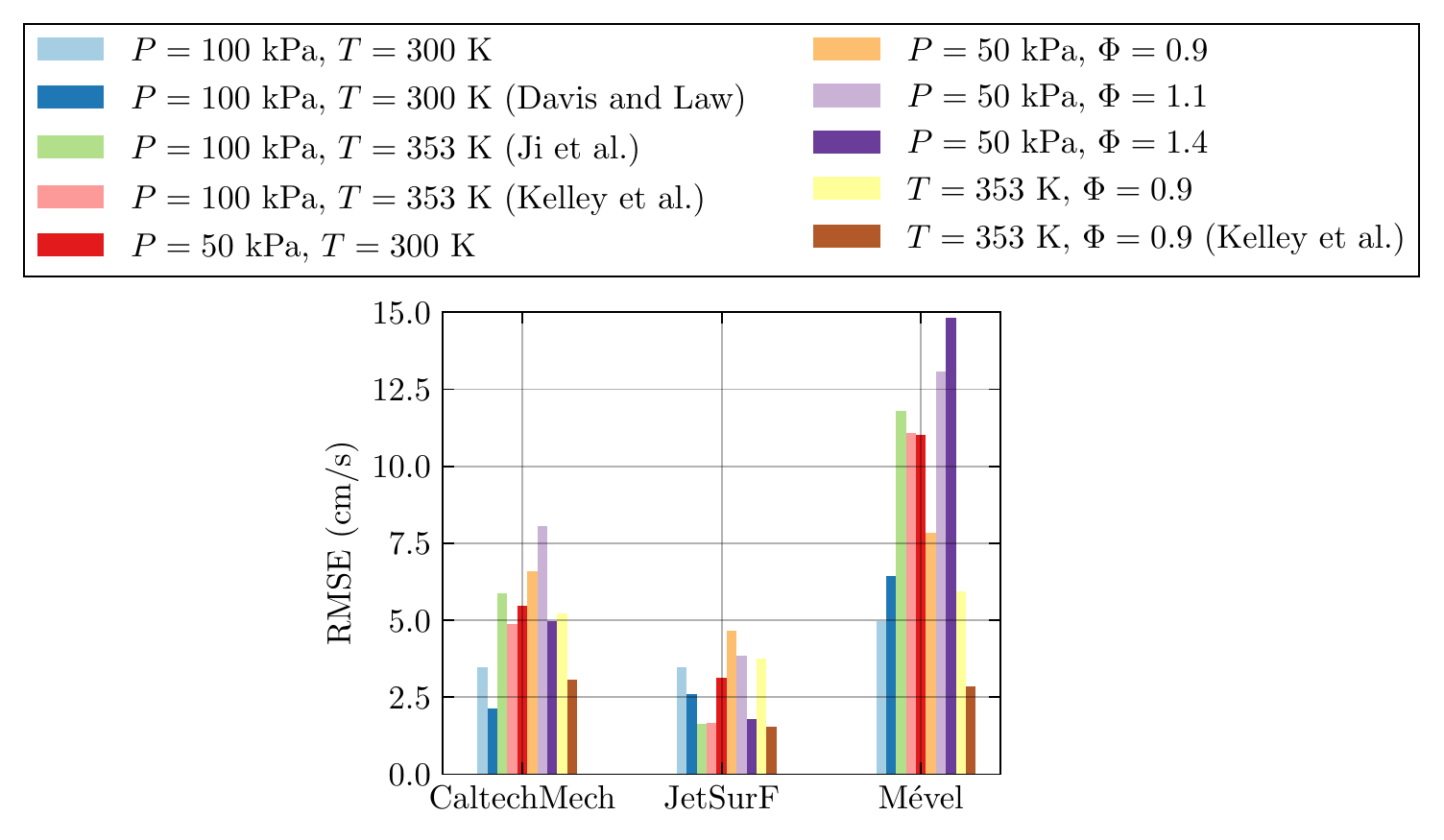}
\caption{Root-mean-squared-error (RMSE) of the calculated laminar burning speeds using CaltechMech \cite{BlanquartMech}, JetSurF \cite{jetsurf}, and M\'evel \cite{MevelFuel2014}.}
\label{fig:RMSE}    
\end{center}
\end{figure}

Figures~\ref{fig:Experimental100kPa} to \ref{fig:ExperimentalT} show comparisons between the experimental and calculated laminar burning speeds. Additional comparisons are shown in Fig.~\ref{fig:T353K} for data from \citet{JiHexaneFlame} and \citet{LawHexane}. Visual inspection of the figures indicates that the chemical kinetic mechanism of M\'evel cannot predict the laminar burning speed with appropriate accuracy. On the other hand, the predictions of CaltechMech and JetSurF  appear to be more accurate; however, it is difficult to ascertain qualitatively which mechanism performs best. The performance of each mechanism is quantitatively evaluated using the root-mean-squared error formulation,
\begin{equation}
\mathrm{RMSE} = \sqrt{\frac{1}{N}\sum\limits_{i=1}^N{\left(S_{\mathrm{calc}}^{(i)}-S_{\mathrm{exp}}^{(i)}\right)^2}},
\end{equation} 
where $S_{\mathrm{calc}}$ and $S_{\mathrm{exp}}$  are the calculated and experimental laminar burning speeds, respectively, $N$ is the number of  points for each experimental data set, and $i$ corresponds to the $i^{\mathrm{th}}$ point in a data set. The RMSE is calculated for the experimental data sets shown in Table~\ref{table:RMSE}. A total of 87 points are used to evaluate the performance of each mechanism, shown in Fig.~\ref{fig:RMSE}.
\begin{table}[ht!]
\begin{center}
\begin{tabular}{cccccc}
\hline
Data & Reference & $P$ (kPa) & $T$ (K) & $\Phi$ & $N$\\
\hline
A & Present study & 100 & 296 & $0.76-1.42$ & 7\\
B & \citet{DavisLaw} & 100 & 300 & $0.85-1.70$ & 16\\
C & \citet{JiHexaneFlame} & 100 & 353 & $0.75-1.50$ & 10\\
D & \citet{LawHexane} & 100 & 353 & $0.75-1.70$ & 19\\
E & Present study & 50 & 296 & $0.86-1.69$ & 12\\
F & Present study & 50 & $297-423$ & 0.9 & 5\\
G &Present study & 50 & $296-422$ & 1.1 & 5\\
H & Present study & 50 & $296-422$ & 1.4 & 5\\
I & Present study & $40-100$ & 357 & 0.9 & 4\\
J & \citet{LawHexane} & $100-1000$ & 353 & 0.9 & 4\\
\hline
\end{tabular}
\caption{Experimental data sets of laminar burning speed used for the RMSE calculation to evaluate the performance of the chemical kinetic mechanisms used in the present study.}
\label{table:RMSE}
\end{center}
\end{table}

Overall, JetSurF yields the smallest RMSE values for almost all the experimental conditions presented in this study and previous studies. The RMSE based on set A  ($P=100$ kPa and $T=300$ K) is the same between JetSurF ($\mathrm{RMSE}=3.5$ cm/s) and CaltechMech; the RMSE based on set B (experiments performed by \citet{DavisLaw}) is smaller, by approximately $19\%$, for CaltechMech ($\mathrm{RMSE}=2.1$ cm/s) than JetSurF ($\mathrm{RMSE}=2.6$ cm/s). For almost all the experimental conditions presented, M\'evel ($\mathrm{RMSE}=2.9-14.8$ cm/s) yields the largest RMSE values when compared to those obtained with JetSurF and CaltechMech. The RMSE based on set J (experiments performed by \citet{LawHexane}) is smaller, by approximately $6\%$, for M\'evel ($\mathrm{RMSE}=2.9$ cm/s) than CaltechMech ($\mathrm{RMSE}=3.1$ cm/s). When considering the RMSE of sets F, G, and H, ($P=50$ kPa and $T\sim300-422$ K) CaltechMech performs best at rich conditions ($\Phi=1.4$); the RMSE  for set H is 5.0 cm/s, approximately 24\% and 38\% smaller than the RMSE  obtained with sets F ($\Phi=0.9$) and G ($\Phi=1.1$), respectively. For JetSurF, set H also has the smallest RMSE (1.8 cm/s) when compared to sets F ($\textrm{RMSE}=4.7$ cm/s) and G ($\textrm{RMSE}=3.9$ cm/s). In regard to the mechanism of M\'evel, the leaner data set F has the smallest RMSE ($7.9$ cm/s) when compared to the close to stoichiometric and rich conditions of sets G ($\textrm{RMSE}=13.1$ cm/s) and H ($\textrm{RMSE}=14.8$ cm/s), respectively. The mean RMSE across the conditions presented in Table~\ref{fig:RMSE} is 5.0 cm/s, 2.8 cm/s, and 9.0 cm/s for CaltechMech, JetSurF, and M\'evel, respectively. Based on a mean RMSE representation of the model performance, JetSurF is the appropriate chemical kinetic mechanism to use when calculating the laminar burning speed of \textit{n}-hexane-air mixtures across a wide range of conditions. The previous statement is made considering the following approach to performing the calculations: a) Soret and Dufour effects were neglected, and b) only mixture-averaged transport properties were considered. 

\subsubsection{Sensitivity Analysis}
\begin{figure}[ht!]
\begin{center}
\begin{tabular}{cc}
\includegraphics[scale=0.7]{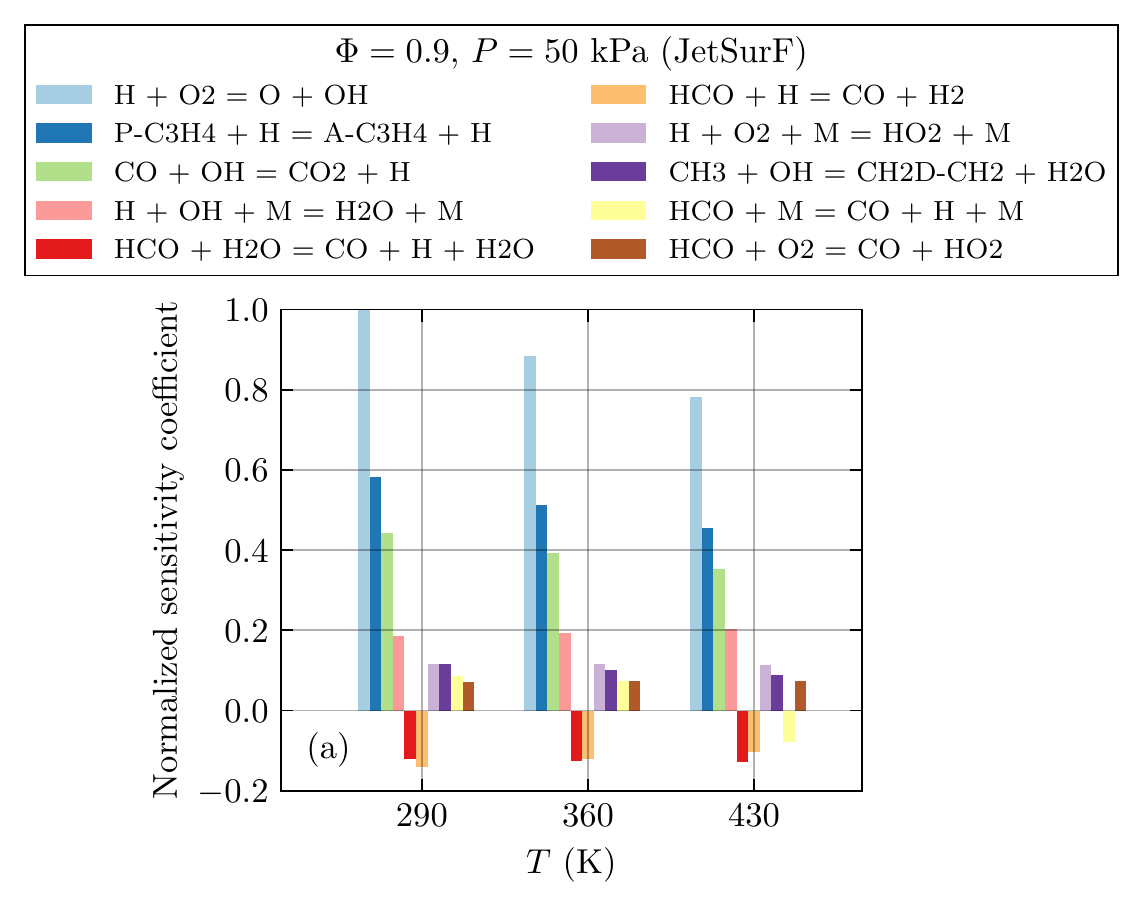} &  \includegraphics[scale=0.7]{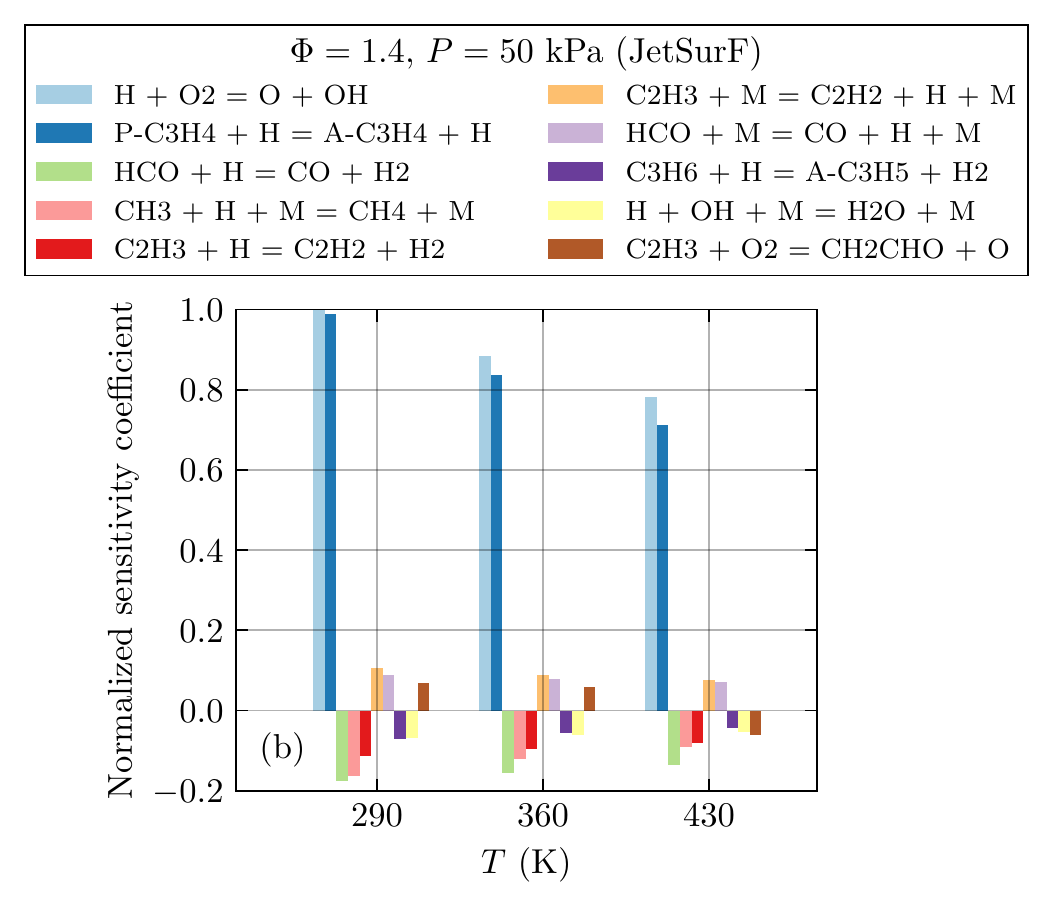}
\end{tabular}
\caption{Normalized sensitivity coefficient as a function of initial temperature at an initial pressure of 50 kPa for (a) $\Phi=0.9$ and (b) $\Phi=1.4$ using JetSurF \citep{jetsurf}.}
\label{fig:SensTemp}
\end{center}
\end{figure}

\begin{figure}[ht!]
\begin{center}
\includegraphics[scale=0.8]{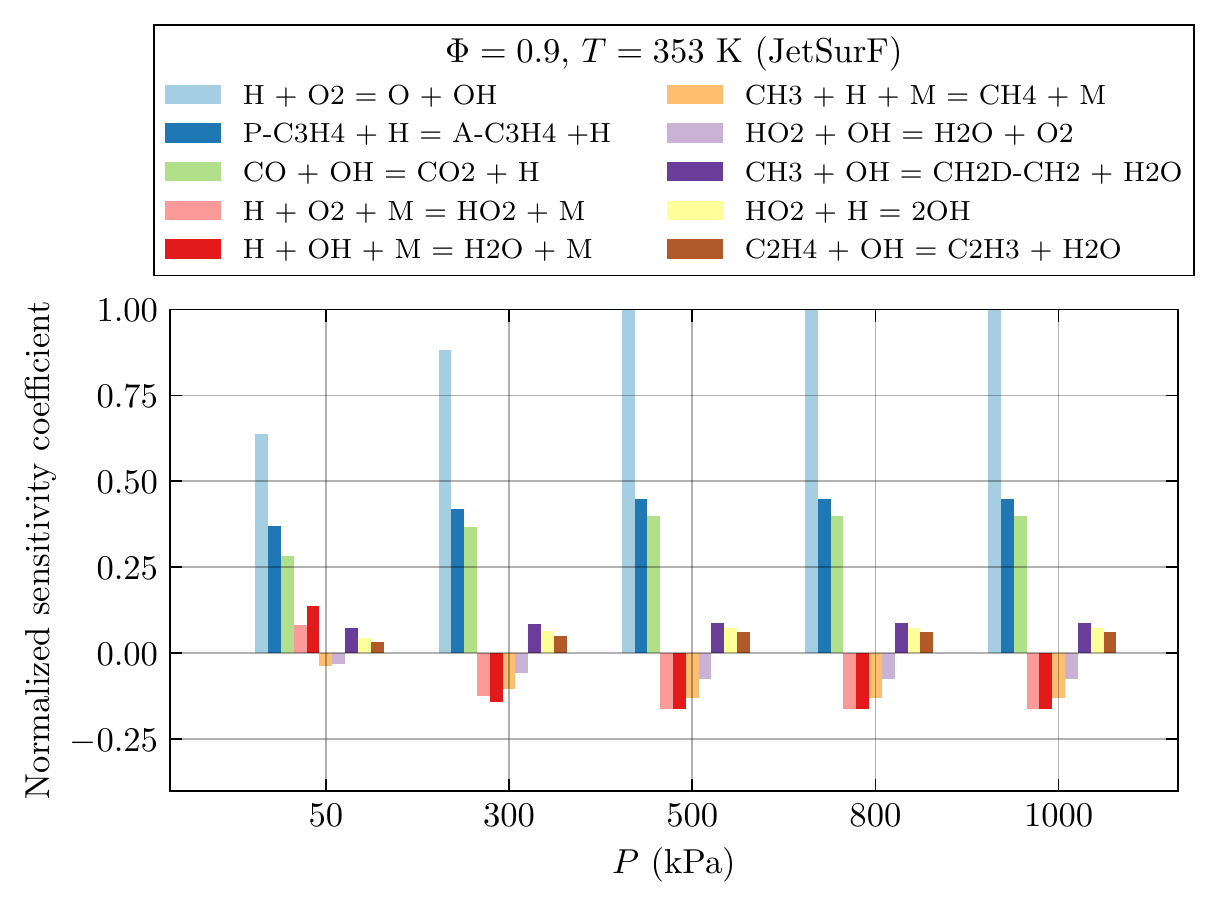}
\caption{Normalized sensitivity coefficient as a function of initial pressure at an initial temperature of 353 K and $\Phi=0.9$ using JetSurF \cite{jetsurf}.}
\label{fig:SensPress}
\end{center}
\end{figure}

A sensitivity analyses was performed with JetSurF to gain further insight into the chemical kinetics of freely propagating \textit{n}-hexane-air flames; the results are shown in Figs.~\ref{fig:SensTemp} and \ref{fig:SensPress}. For all the conditions tested, the most important reaction was the chain-branching reaction R$_{1}$: H+O$_{2}$=OH+O. The sensitivity coefficient of this reaction increases as pressure increases and decreases as temperature increases. The second most sensitive reaction for all conditions tested was R$_{2}$: p-C$_{3}$H$_{4}$+H=A-C$_{3}$H$_{4}$+H which exhibited a positive coefficient. For the lean mixture ($\Phi=0.9$), the third most important reaction for all temperatures and pressures investigated was R$_{3}$: CO+OH=CO$_{2}$+H. R$_3$ is important due to: (1) it's high exothermicity which contributes to a temperature increase and speeds up the overall reaction rate, and (2) the generation of the H atom. The fourth most important reaction for the lean mixture was the recombination reaction R$_{4}$: H+OH(+M)=H$_2$O(+M). At low pressure, and for all the temperatures tested, the sensitivity coefficient of R$_{4}$ was positive. However, as the pressure increased, the sensitivity coefficient became negative. This is due to the increased competition between the chain branching reaction R$_1$ and the termination reaction R$_4$ as pressure increases.

For the rich mixture ($\Phi=1.4$), as a result of the deficiency of oxygen, reactions R$_{3}$ and R$_{4}$ do not appear within the most important reactions. The reactions R$_{5}$: HCO+H=CO+H$_{2}$ and R$_{6}$: CH$_{3}$+H(+M)=CH$_{4}$+H(+M) exhibited negative sensitivity coefficients because they reduce the pool of free radicals by consuming the H atom. 

\subsubsection{Reactions Pathway Analysis}

\begin{figure}[!h]
\begin{center}
\includegraphics[width=0.7\linewidth]{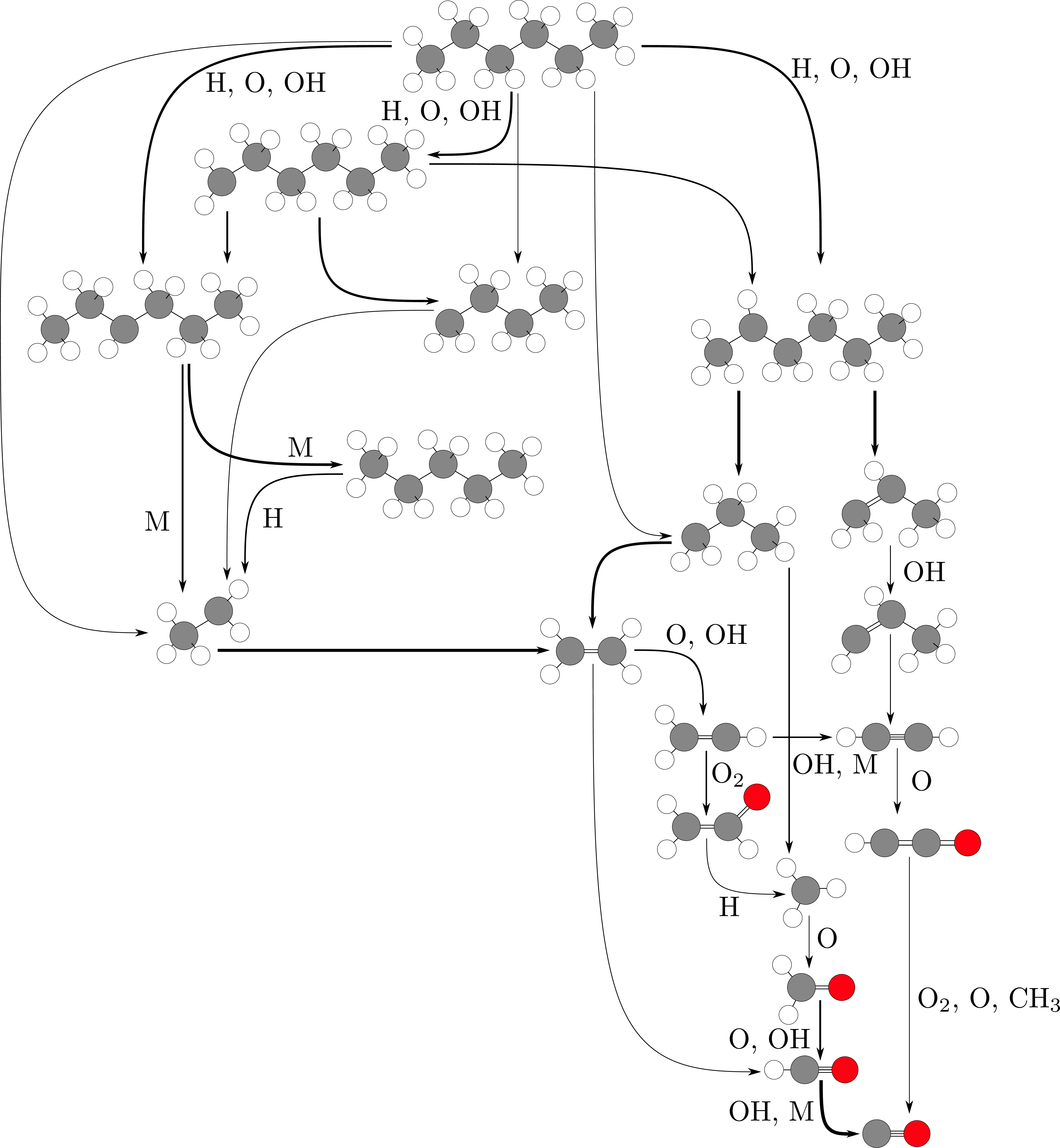} 
\caption{Reaction pathway analysis for the carbon element in a lean \textit{n}-hexane-air flame at $\Phi$ = 0.90 and initial temperature and initial pressure of 296 K and 50 kPa, respectively. Position and temperature in the flame are 4.9 mm and 1443 K, respectively.}
\protect\label{Pathways}      
\end{center}
\end{figure}

A reaction pathway analysis was performed using Cantera \cite{goodwin} for a lean \textit{n}-hexane-air mixture at $\Phi=0.90$ and initial temperature and initial pressure of 296 K and 50 kPa, respectively, using JetSurF. The reaction pathway was obtained as elementary mass fluxes and was performed with a threshold of 10\% in order to focus on the most important pathways. Figure~\ref{Pathways} shows a typical example of a reaction pathway obtained at a distance of 4.9 mm from the flame front and a corresponding temperature of 1443 K. Hexane consumption is mainly driven by H-abstraction reactions, with the OH radical being the most efficient abstracter. The 1-hexyl radical undergoes isomerization which increases the yields of 2-hexyl and 3-hexyl radicals. Conversely, hexane undergoes C-C bond fission leading to ethyl, propyl and butyl radicals. The consumption of 2-hexyl and 3-hexyl radicals also occurs mainly through C-C bond rupture which leads to the formation of a significant amount of C$_2$H$_4$. Ethylene consumption eventually leads to CO formation mainly though the following sequences:

\begin{equation}
\text{C}_2\text{H}_4 \xrightarrow{\text{OH}} \text{C}_2\text{H}_3 \xrightarrow{\text{OH}} \text{C}_2\text{H}_2 \xrightarrow{\text{O}} \text{HCCO} \xrightarrow{\text{O}_2} \text{CO}  
\label{FormationCO1}
\end{equation}
and
\begin{equation}
\text{C}_2\text{H}_4 \xrightarrow{\text{OH}} \text{C}_2\text{H}_3 \xrightarrow{\text{O}_2} \text{CH}_2\text{CO} \xrightarrow{\text{H}} \text{CH}_3 \xrightarrow{\text{O}} \text{CH}_2\text{O} \xrightarrow{\text{OH}} \text{HCO} \xrightarrow{\text{OH}} \text{CO}  .
\label{FormationCO2}
\end{equation}

At the temperature considered, no significant conversion of CO into CO$_2$ was detected. This reaction pathway analysis underlines the importance of ethylene which appears as a ``bottle-neck" species in the course of hexane oxidation.
\section{Summary}
\textit{n}-Hexane-air mixtures were characterized through experimental measurements and calculations of the laminar burning speed. The laminar burning speed was obtained by using a nonlinear methodology. The effect of equivalence ratio, temperature, and pressure on the laminar burning speed was investigated experimentally by varying the equivalence ratio $\Phi=0.62-1.60$, the initial temperature from $296$ K to $422$ K, and the initial pressure from $50$ kPa to $100$ kPa. The laminar burning speed was observed to increase as pressure decreases ($T=357$ K) and as temperature increases. It was also shown that the laminar burning speed increases at comparable rates as temperature increases for mixtures at $\Phi=\{0.90,1.10,1.40\}$. The predictive capabilities of three chemical kinetic mechanisms from the literature were quantitatively evaluated using the present experimental data and those from the literature. Based on a RMSE analysis, it was shown that JetSurF was the most appropriate mechanism for modeling the laminar burning speed of \textit{n}-hexane-air mixtures over a wide range of mixture compositions and thermodynamic conditions.

\section*{Acknowledgments}

This work was carried out in the Explosion Dynamics Laboratory of the California Institute of Technology, and was supported by The Boeing Company through a Strategic Research and Development Relationship Agreement CT-BA-GTA-1.

\section*{References}
\bibliographystyle{elsarticle-num-CNF}
\bibliography{bib/Biofuels,bib/chemistry,bib/Nonlinear,bib/EDLpapers,bib/flames,bib/Laminar,bib/ChemicalModels,bib/software,bib/EDLpapers,bib/MESDOC}
\clearpage
\newpage
\appendix
\section{Statistical Analysis: Mann-Whitney-Wilcoxon (MWW) RankSum Test}
The Mann-Whitney-Wilcoxon (MWW) RankSum test was used to determine if the distribution of measurements in set $\mathbf{A}$ were the same as the results from set $\mathbf{B}$, written symbolically as the null hypothesis $H_0: \mathbf{A} =\mathbf{B}$.  The test also detects shifts in the distributions given by sets $\mathbf{A}$ and $\mathbf{B}$, written as the hypothesis $H_1 : \mathbf{A} \ne \mathbf{B}$. The test ranks $n_{\mathbf{A}}+n_{\mathbf{B}}$ observations of the combined distributions, where $n_{\mathbf{A}}$ and $n_{\mathbf{B}}$ correspond to the number of experimental observations in sets  $\mathbf{A}$ and $\mathbf{B}$, respectively. Each observation has a rank, where rank 1 and rank $n_{\mathbf{A}}+n_{\mathbf{B}}$ correspond to the smallest and largest values of $S_u^0$. In the following example, set $\mathbf{A}$ and set $\mathbf{B}$  correspond to Data A and Data B, respectively, from Table~\ref{table:RMSE}.  The sum of the rank of set $\mathbf{B}$ is $W=200$; under the null hypothesis $H_0$, the mean and variance of $W$ is,
\begin{equation}
\mu_W = \frac{n_{\mathbf{B}}(N+1)}{2}=\frac{16(23+1)}{2}=192,
\end{equation}
\begin{equation}
\sigma_W^2 = \frac{n_{\mathbf{A}}n_{\mathbf{B}}(N + 1)}{12}=\frac{7\cdot 16(23+1)}{12}=224,
\end{equation}
where $N=n_{\mathbf{A}}+n_{\mathbf{B}}$. The observed value of the test statistic is,
\begin{equation}
Z_{\mathrm{obs}}=\frac{W-\mu_W}{\sigma_W}=\frac{200-192}{\sqrt{224}}=0.5
\end{equation}
The two-tailed p-value (calculated probability), p, at $Z_{\mathrm{obs}}=0.5$ is,
\begin{equation}
\mathrm{p}(0.5) = 2\left(1-\int\limits_{-\infty}^{0.5}\frac{1}{\sqrt{2\pi}}e^{-z^2/2}\mathrm{d}z\right)=0.6
\end{equation}
Since $\mathrm{p}>0.05$ the differences between sets $\mathbf{A}$ and $\mathbf{B}$ are not statistically significant. 

The MWW RankSum test was used to compare the laminar burning speeds from Data A and Data E, at 100 kPa (set $\mathbf{A}$) and 50 kPa (set $\mathbf{B}$) respectively. The sum of the rank of set $\mathbf{B}$ is $123$; under the null hypothesis $H_0$, the mean and variance of $W$ is 120 and 140, respectively. The calculated $Z_{\mathrm{obs}}$ is 0.3 resulting in a p-value of 0.8;  since $\mathrm{p}>0.05$, the differences between sets $\mathbf{A}$ and $\mathbf{B}$ are not statistically significant. 
\newpage
\section{Present Study Experimental Results}
\setcounter{table}{0}
\begin{table}[h!]
  \begin{center}
    \begin{filecontents*}{Caltech1.csv}
Number,Phi,T1(K),P1(kPa),Lb(cm),Uncertainty,Su,uncertainty
0,1.06,296,100,0.042,0.005,36,2
1,0.76,291,100,0.208,0.019,24,1
2,0.83,296,100,0.153,0.015,29,2
3,0.90,296,100,0.117,0.013,33,2
4,0.98,296,100,0.115,0.012,35,2
5,1.26,296,100,0.067,0.007,36,2
6,1.43,297,100,-0.028,0.006,28,2
9,1.30,298,50,0.041,0.006,39,2
18,1.00,296,50,0.143,0.016,38,2
20,0.89,297,50,0.215,0.018,33,2
24,0.85,297,50,0.187,0.017,35,2
26,0.99,295,50,0.181,0.017,35,2
27,1.10,295,50,0.132,0.018,40,2
29,1.20,296,50,0.098,0.012,39,2
30,1.20,296,50,0.099,0.013,39,2
31,1.30,296,50,0.034,0.004,34,2
32,1.40,296,50,-0.070,0.013,28,2
33,1.50,296,50,-0.300,0.042,20,1
34,1.58,297,50,-0.531,0.031,13,1
38,1.00,297,50,0.152,0.017,39,2
39,1.11,297,50,0.123,0.014,40,2
40,0.90,297,50,0.183,0.017,36,2
41,1.34,297,50,0.001,0.005,31,2
42,1.69,297,50,-1.097,0.087,9,1
\end{filecontents*}
\pgfplotstabletypeset[
      multicolumn names, 
      col sep=comma, 
      trim cells=true,
      display columns/0/.style={
		column name=Test, 
		column type={c},string type,fixed zerofill,precision=0,dec sep align},
		display columns/1/.style={
		column name=$\Phi$,
		column type={S},string type},
		display columns/2/.style={
		column name=$T_0$ (K),
		column type={S},string type,fixed zerofill,precision=0,dec sep align},
		display columns/3/.style={
		column name=$P_0$ (K),
		column type={c},string type,fixed zerofill,precision=0,dec sep align},
		display columns/4/.style={
		column name=$L_B$ (cm),
		column type={S},string type},
		display columns/5/.style={
		column name=$\Delta L_B$ (cm),
		column type={S},string type},
		display columns/6/.style={
		column name=$S_u^0$ (cm/s),
		column type={S},string type},
		display columns/7/.style={
		column name=$\Delta S_u^0$ (cm/s),
		column type={c},string type},
      every head row/.style={
		before row={\midrule}, 
		after row={\midrule} 
			},
		every last row/.style={after row=\midrule}, 
    ]{Caltech1.csv} 
  \end{center}
    \caption{Results of spherically expanding flame experiments performed at Caltech.}
    \label{table:Caltech1}
\end{table}

\begin{table}[h!]
  \begin{center}
    \begin{filecontents*}{Caltech2.csv}
Number,Phi,T1(K),P1(kPa),Lb(cm),Uncertainty,Su,Uncertainty
43,0.95,297,50,0.154,0.015,35,2
44,0.86,297,50,0.177,0.016,34,2
45,0.92,313,50,0.173,0.018,40,2
46,1.11,312,50,0.113,0.013,43,3
47,1.41,313,50,-0.057,0.009,29,2
48,0.89,314,50,0.184,0.018,38,2
49,0.90,326,50,0.168,0.026,40,2
50,1.10,325,50,0.138,0.015,46,3
51,1.40,325,50,0.002,0.003,33,2
52,0.90,357,50,0.171,0.015,48,3
53,1.10,358,50,0.120,0.014,53,3
54,1.40,358,50,-0.016,0.003,37,2
55,0.90,380,50,0.162,0.016,52,3
56,1.10,380,50,0.117,0.014,58,3
57,1.40,380,50,0.003,0.001,43,3
58,0.90,357,100,0.080,0.010,42,3
59,0.89,357,80,0.106,0.013,42,3
61,0.90,357,40,0.177,0.017,47,3
\end{filecontents*}
\pgfplotstabletypeset[
      multicolumn names, 
      col sep=comma, 
      trim cells=true,
      display columns/0/.style={
		column name=Test, 
		column type={c},string type,fixed zerofill,precision=0,dec sep align},
		display columns/1/.style={
		column name=$\Phi$,
		column type={S},string type},
		display columns/2/.style={
		column name=$T_0$ (K),
		column type={S},string type,fixed zerofill,precision=0,dec sep align},
		display columns/3/.style={
		column name=$P_0$ (K),
		column type={c},string type,fixed zerofill,precision=0,dec sep align},
		display columns/4/.style={
		column name=$L_B$ (cm),
		column type={S},string type},
		display columns/5/.style={
		column name=$\Delta L_B$ (cm),
		column type={S},string type},
		display columns/6/.style={
		column name=$S_u^0$ (cm/s),
		column type={S},string type},
		display columns/7/.style={
		column name=$\Delta S_u^0$ (cm/s),
		column type={c},string type},
      every head row/.style={
		before row={\midrule}, 
		after row={\midrule} 
			},
		every last row/.style={after row=\midrule}, 
    ]{Caltech2.csv} 
  \end{center}
    \caption{Results of spherically expanding flame experiments performed at Caltech [continued].}
    \label{table:Caltech2}
\end{table}

\begin{table}[h!]
  \begin{center}
    \begin{filecontents*}{ICARE.csv}
Equivalence Ratio,Initial T (K),Pressure (kPa),SL (cm/s),Uncertainty
1.51,424,50,37,2
1.40,422,50,53,3
1.31,422,50,64,4
1.21,423,50,71,4
1.12,423,51,73,4
1.05,423,51,73,4
0.99,423,51,71,4
0.89,423,51,67,4
1.02,425,51,74,4
1.17,425,51,73,4
\end{filecontents*}
\pgfplotstabletypeset[
      multicolumn names, 
      col sep=comma, 
      trim cells=true,
		display columns/0/.style={
		column name=$\Phi$,
		column type={S},string type},
		display columns/1/.style={
		column name=$T_0$ (K),
		column type={S},string type,fixed zerofill,precision=0,dec sep align},
		display columns/2/.style={
		column name=$P_0$ (K),
		column type={c},string type,fixed zerofill,precision=0,dec sep align},
		display columns/3/.style={
		column name=$S_u^0$ (cm/s),
		column type={S},string type},
		display columns/4/.style={
		column name=$\Delta S_u^0$ (cm/s),
		column type={c},string type},
      every head row/.style={
		before row={\midrule}, 
		after row={\midrule} 
			},
		every last row/.style={after row=\midrule}, 
    ]{ICARE.csv} 
  \end{center}
    \caption{Results of spherically expanding flame experiments performed at ICARE-CNRS.}
    \label{table:ICARE}
\end{table}
\clearpage
\newpage
\section{Previous Work Experimental Results}
\setcounter{table}{0}

\begin{table}[h!]
  \begin{center}
    \begin{filecontents*}{davis.csv}
Phi,Su
0.80,27
0.85,31
0.90,35
0.95,37
1.00,39
1.05,40
1.10,40
1.15,40
1.20,38
1.25,36
1.30,33
1.35,29
1.40,26
1.50,18
1.60,14
1.70,12
\end{filecontents*}
\pgfplotstabletypeset[
      multicolumn names, 
      col sep=comma, 
      skip first n={1},
      display columns/0/.style={
		column name=$\Phi$, 
		column type={S},string type},  
      display columns/1/.style={
		column name=$S_u^0$ (cm/s),
		column type={S},string type},
      every head row/.style={
		before row={\midrule}, 
		after row={\midrule} 
			},
		every last row/.style={after row=\midrule}, 
    ]{davis.csv} 
  \end{center}
       \caption{Results of experimental laminar burning speeds obtained by \citet{DavisLaw} at a nominal initial temperature and pressure of 300 K and 100 kPa, respectively.}
    \label{table:davis}
\end{table}

\begin{table}[h!]
  \begin{center}
    \begin{filecontents*}{kelley.csv}
P,Su
100,42
200,37
500,30
1000,25
\end{filecontents*}
\pgfplotstabletypeset[
      multicolumn names, 
      col sep=comma, 
      skip first n={1},
      display columns/0/.style={
		column name=$P_0$ (kPa), 
		column type={S},string type},  
      display columns/1/.style={
		column name=$S_u^0$ (cm/s),
		column type={S},string type},
      every head row/.style={
		before row={\midrule}, 
		after row={\midrule} 
			},
		every last row/.style={after row=\midrule}, 
    ]{kelley.csv} 
  \end{center}
       \caption{Results of experimental laminar burning speeds obtained by \citet{LawHexane} at an equivalence ratio and nominal initial temperature of 0.9 and 353 K, respectively.}
    \label{table:kelleyPress}
\end{table}

\begin{table}[h!]
  \begin{center}
    \begin{filecontents*}{Farrel.csv}
Phi,SL (cm/s)
0.55,19
0.60,25
0.70,41
0.80,52
0.90,59
1.00,70
1.10,74
1.20,76
1.30,72
\end{filecontents*}
\pgfplotstabletypeset[
      multicolumn names, 
      col sep=comma, 
      display columns/0/.style={
		column name=$\Phi$, 
		column type={S},string type},  
      display columns/1/.style={
		column name=$S_u^0$ (cm/s),
		column type={c},string type,precision=0},
      every head row/.style={
		before row={\midrule}, 
		after row={\midrule} 
			},
		every last row/.style={after row=\midrule}, 
    ]{Farrel.csv} 
  \end{center}
     \caption{Results of experimental laminar burning speeds obtained by \citet{FarellHexane} at a nominal initial temperature and pressure of 450 K and 304 kPa, respectively.}
    \label{table:farrel}
\end{table}

\begin{table}[h!]
  \begin{center}
    \begin{filecontents*}{law.csv}
Phi,Su
0.75,31
0.80,35
0.85,40
0.90,42
0.95,45
1.00,46
1.05,48
1.10,48
1.15,48
1.20,46
1.25,44
1.30,40
1.35,35
1.40,31
1.45,28
1.50,24
1.55,20
1.60,17
1.70,11
\end{filecontents*}
\pgfplotstabletypeset[
      multicolumn names, 
      col sep=comma, 
      display columns/0/.style={
		column name=$\Phi$, 
		column type={S},string type},  
      display columns/1/.style={
		column name=$S_u^0$ (cm/s),
		column type={S},string type},
      every head row/.style={
		before row={\midrule}, 
		after row={\midrule} 
			},
		every last row/.style={after row=\midrule}, 
    ]{law.csv} 
  \end{center}
     \caption{Results of experimental laminar burning speeds obtained by \citet{LawHexane} at a nominal initial temperature and pressure of 353 K and 100 kPa, respectively.}
    \label{table:kelley}
\end{table}

\begin{table}[h!]
  \begin{center}
    \begin{filecontents*}{fokion.csv}
Phi,Su
0.75,37
0.80,40
0.95,48
1.00,50
1.05,50
1.10,49
1.20,45
1.30,38
1.40,27
1.50,19
\end{filecontents*}
\pgfplotstabletypeset[
      multicolumn names, 
      col sep=comma, 
      skip first n={1},
      display columns/0/.style={
		column name=$\Phi$, 
		column type={S},string type},  
      display columns/1/.style={
		column name=$S_u^0$ (cm/s),
		column type={S},string type},
      every head row/.style={
		before row={\midrule}, 
		after row={\midrule} 
			},
		every last row/.style={after row=\midrule}, 
    ]{fokion.csv} 
  \end{center}
       \caption{Results of experimental laminar burning speeds obtained by \citet{JiHexaneFlame} at a nominal initial temperature and pressure of 353 K and 100 kPa, respectively.}
    \label{table:ji}
\end{table}
\end{document}